\documentstyle[seceq,preprint]{ptptex}
\newcommand{\bra}[1]{\langle {#1} |}     %%
\newcommand{\ket}[1]{| {#1} \rangle}     %%
\newcommand{\kket}[1]{| {#1} \rangle\!\rangle}     %%
\title{%        %You can use \\ for explicit line-break
A Possible Form of the Orthogonal Set\\
in Six Kinds of Boson Operators
}
\subtitle{
In Relation to the su(2)- and its Relevant Algebras
}    %use this when you want a subtitle

\author{%       %Use \sc for the family name
Atsushi {\sc Kuriyama},$^{1}$ 
Constan\c{c}a {\sc Provid\^encia},$^{2}$ \\
Jo\~ao da {\sc Provid\^encia},$^{2}$ Yasuhiko {\sc Tsue}$^{3}$ 
and Masatoshi {\sc Yamamura}$^{1}$
}

\inst{%         %Affiliation, neglected when [addenda] or [errata]
$^1$Faculty of Engineering, Kansai University, Suita 564-8680, Japan\\
$^{2}$Departamento de Fisica, Universidade de Coimbra, P-3000 Coimbra, 
Portugal\\
$^{3}$Physics Division, Faculty of Science, Kochi University, Kochi 780-8520, 
Japan
}

\recdate{%      %Editorial Office will fill in this.
}

\abst{%       %this abstract is neglected when [addenda] or [errata]
With the aim of constructing coherent states for many-body systems 
consisting of six kinds of boson operators, a possible form of the orthogonal 
set is presented in terms of monomial with respect to 
state generating operators. In connection with the $su(3)$-, the $su(2)$-, 
the $su(2,1)$- and the $su(1,1)$-algebras, four types of the orthogonal 
set are discussed.}

\begin{document}

\maketitle

\section{Introduction}

Time dependent variational method has played a central role for 
describing time-evolution of quantal many-body systems. 
The typical example may be the time-dependent Hartree-Fock (TDHF) 
method for many-fermion systems. In this case, a trial state for the 
variation is the Slater determinant containing complex parameters. 
Especially, it may be interesting to see that, under the time-dependent 
variation with a certain condition governing these parameters, 
time-evolution of many-fermion systems under investigation 
can be treated in the framework of classical mechanics in 
Hamilton form.\cite{1,2} 
In the case of many-boson systems, the time-evolution of the systems 
can be also described in the framework of the time-dependent variational 
method. In this case, the simplest trial state may be the boson coherent 
state for one kind of boson operator. 
However, in order to give the variety to the variation, the trial state 
must be modified from the boson coherent state to other states. 
For a possible modification, we know that with the help of a function 
of the boson number operator, the boson coherent state can be generalized. 
This generalization may be called the deformation of the boson 
coherent state and Penson and Solomon maintained that the choice of 
the function characterizing the deformation is rather arbitrary.\cite{3} 
By adopting various forms of the functions, various trial states are 
obtained. In response to their opinion, the present authors proposed 
the deformed boson scheme for a generalizing boson coherent state. 
In this case, also, we can formulate classical mechanics in Hamiltonian 
form under the same idea as that in the TDHF method.\cite{4}

In order to approach to various realistic problems as much as possible, 
it is necessary to formulate boson coherent states for many-body systems 
consisting of many kinds of boson operators and, further, we make the 
boson coherent states deform. In order to construct such boson 
coherent states, it is inevitable to prepare the orthogonal set based on the 
algebraic examination as a preliminary task. 
Typical examples of this task can be found in the $su(2)$- and the 
$su(1,1)$-algebras. The generators are expressed in terms of bilinear forms 
for two kinds of bosons. This is called the Schwinger boson 
representation.\cite{5} The case of three kinds of bosons was formulated 
by the present authors in relation to the $su(2,1)$-algebra.\cite{6} 
In addition to the above cases, the present authors formulated the case 
of four kinds of bosons, in which the $su(2)\times su(2)$- and the 
$su(1,1)\times su(1,1)$-algebras are objects of the investigation.\cite{7} 
This is a simple example of the formalism in which $(M+1)(N+1)$ kinds 
of bosons are treated.\cite{8} 
In each case, the orthogonal set was investigated. As the first step 
for constructing the boson coherent state after finishing the algebraic 
examination, we set up an exponential type superposition of various states 
with the minimum weight or its equivalent state. 
At the second step, we set up an operator, the form of which is of 
exponential superposition of various raising or coherent state generating 
operators. As the final step, by operating the exponential operator 
obtained at the second step on the state introduced at the first step, 
we are able to obtain the coherent state. 
Once the coherent state is given, the task for the deformation may be 
not so tedious as that imagined. 
With the aid of the deformed boson scheme, it is completed. 
This deformation leads us to the concept of the $q$-deformed algebra. 
Of course, in each case, each device is necessary. 
Following the above-mentioned viewpoint, the present authors investigated 
the cases of two,\cite{9} three\cite{10} and four\cite{11} kinds of 
boson operators.

Main aim of this paper is to show a possible orthogonal set for six kinds 
of boson operators. There exist two reasons why we investigate the above case. 
In Ref. \citen{12}, we proposed a method for describing statistically 
mixed state of many-body systems consisting of one kind of boson, which 
interacts with an external harmonic oscillator. The basic idea partly comes 
from the phase space doubling which characterized the thermo field 
dynamics formalism.\cite{13} 
And, in Ref. \citen{10}, this method is transparently understood in the 
framework of the deformation of the $su(2,1)$-algebra in three kinds 
of bosons. Therefore, as a natural extension, we have an idea how to describe 
statistically mixed state of many-body systems consisting of two kinds of 
bosons, which interact with an external harmonic oscillator. 
Such many-body systems are found in the single-level shell model under 
the pairing correlation and the two-level shell model under the 
particle-hole interaction, i.e., Lipkin model. 
Through this investigation, we are able to have an understanding of 
a possible mechanism of damping or amplifying of the collective motions 
which are observed in these models. 
This is the first reason. 
The second reason is related to the formalism given in Ref. \citen{11}. 
In this work, with the help of the deformed boson scheme, we proposed a method 
which enable us to describe thermal effects observed in the two-level 
shell model under the pairing correlation. 
Therefore, it may be interesting to check if the basic idea presented in 
Ref. \citen{11} is applicable or not for the three-level shell model. 
This is the second reason. The above-mentioned two reasons suggest us the 
importance of the investigation of the case of six kinds of boson operators. 
For this aim, first, we must present an orthogonal set which is convenient 
for constructing the coherent state.

Our starting point can be found in the paper by the present authors which 
is referred to as Ref. \citen{8}. 
In this paper, we can construct the $su(M+1)\times su(N,1)$-algebra in the 
framework of $(M+1)(N+1)$ kinds of boson operators. 
In the present paper, we treat the case $M=1$ and $N=2$. Then, in this case, 
totally six kinds of bosons are treated and the $su(2)$- and 
the $su(2,1)$-algebras are obtained. Further, in this framework, we can 
construct the $su(1,1)$- and the $su(3)$-algebras. 
Since the system under investigation consists of six kinds of bosons, 
the orthogonal set is specified by six quantum numbers. 
Therefore, in order to get the orthogonal set, it may be necessary to 
prepare six mutually commutable hermitian operators. 
In fact, as will be shown in \S\S 4 and 5, we are able to prepare four sets 
of these operators and, then, four types of the orthogonal sets exist. 
These four sets are related to the $su(3)$-, the $su(2)$-, the $su(2,1)$- 
and the $su(1,1)$-algebras, respectively. Further, the orthogonal set can be 
expressed in terms of monomial as a function of the state generating 
operators. This fact is quite important for constructing the coherent state. 
Therefore, we can expect that the orthogonal set obtained in this paper 
plays a powerful role for constructing the coherent state 
consisting of six kinds of bosons.

In the next section, the basic framework for investigating the orthogonal set 
is summarized. In \S 3, a possible form of the orthogonal set is presented 
in terms of polynomial as a function of state generating operators. 
Sections 4 and 5 are central parts of this paper. The orthogonal set is given 
in terms of monomial for four types related to the $su(3)$-, the $su(2)$-, 
the $su(2,1)$- and the $su(1,1)$-algebras. Finally, in \S 6, two points are 
discussed as concluding remarks : 
1) the relation of the formalism in this paper with the $su(2,1)$-model 
in three kinds of bosons and 2) a basic idea for construction of the 
coherent state.

\section{The su(2)-algebra in six kinds of boson operators}

In this section, we recapitulate the $su(2)$-algebra in six kinds of 
boson operators in a form slightly different from that developed 
by the present authors in Ref. \citen{8} including the notations. 
Let us denote six kinds of bosons $({\hat a}_i , {\hat a}_i^*)$ 
and $({\hat b}_i , {\hat b}_i^*)$ ($i=1,2,3$). 
With the use of these bosons, the following operators obeying the 
$su(2)$-algebra are defined : 
\begin{equation}\label{2-1}
{\hat S}_{\pm, 0}=\sum_{i=1}^3 {\hat S}_{\pm, 0}(i) \ , 
\end{equation}
\begin{subequations}\label{2-2}
\begin{eqnarray}
& &{\hat S}_+(i)={\hat a}_i^*{\hat b}_i \ , \qquad 
{\hat S}_-(i)={\hat b}_i^*{\hat a}_i \ , \nonumber\\
& &{\hat S}_0(i)=(1/2)\cdot ({\hat a}_i^*{\hat a}_i -{\hat b}_i^*{\hat b}_i) 
\ . \quad (i=1,2,3)
\label{2-2a}
\end{eqnarray}
The above corresponds to the case $(M=1, N=2)$ in Ref. \citen{8}. 
Associating with the operators (\ref{2-2a}), we introduce the operators 
defined as 
\begin{equation}\label{2-2b}
{\hat S}(i)=(1/2)\cdot ({\hat a}_i^*{\hat a}_i+{\hat b}_i^*{\hat b}_i) \ .
\quad (i=1,2,3)
\end{equation}
\end{subequations}
Three sets $({\hat S}_{\pm,0}(i) ; i=1,2,3)$ form mutually independent 
$su(2)$-algebra in the Schwinger boson representation\cite{5} and the 
operator ${\hat S}(i)$ denotes the magnitude of the $i$-th $su(2)$-spin. 
Therefore, $({\hat S}_{\pm,0})$ forms the addition of these three 
$su(2)$-spins. First aim of this paper is to give the eigenstate of 
$({\hat {\mib S}}^2 , {\hat S}_0)$ without the limits of the conventional 
manner, i.e., successive addition of the $su(2)$-spins. 
Here, ${\hat {\mib S}}^2$ denotes the Casimir operator of the $su(2)$-spin : 
\begin{equation}\label{2-3}
{\hat {\mib S}}^2={\hat S}_0^2+(1/2)\cdot 
({\hat S}_+{\hat S}_- + {\hat S}_-{\hat S}_+)
={\hat S}_0({\hat S}_0-1)+{\hat S}_+{\hat S}_- \ . 
\end{equation}
The eigenstate of $({\hat {\mib S}}^2 , {\hat S}_0)$ with the eigenvalues 
$(s(s+1) , s_0)$ satisfies 
\begin{eqnarray}
& &{\hat {\mib S}}^2\ket{(\alpha) ; s s_0}=s(s+1)\ket{(\alpha) ; s s_0} \ , 
\nonumber\\
& &{\hat S}_0\ket{(\alpha) ; s s_0}=s_0\ket{(\alpha) ; s s_0} \ . 
\label{2-4}\\
& &\quad s=0, 1/2, 1, 3/2, \cdots , \qquad 
s_0=-s, -s+1, \cdots , s-1, s \ .
\label{2-5}
\end{eqnarray}
Here, $(\alpha)$ denotes a set of four numbers additional to 
$(s, s_0)$ which specifies the eigenstate of only $({\hat {\mib S}}^2 , 
{\hat S}_0)$. It should be noted that the four numbers do not play 
a role of quantum numbers directly. 
The state $\ket{(\alpha) ; s s_0}$ can be expressed in the form 
\begin{eqnarray}
& &\ket{(\alpha) ; s s_0}=\left(\sqrt{(s+s_0)!}\right)^{-1}
({\hat S}_+)^{s+s_0}\ket{(\alpha) ; s} \ , 
\label{2-6}\\
& &{\hat S}_-\ket{(\alpha) ; s}=0 \ , \qquad
{\hat S}_0\ket{(\alpha) ; s }=-s\ket{(\alpha) ; s} \ . 
\label{2-7}
\end{eqnarray}
We call the state $\ket{(\alpha) ; s}$  the intrinsic state, 
because $\ket{(\alpha) ; s s_0}$ is constructed by successive operation of 
${\hat S}_+$ on the state $\ket{(\alpha) ; s}$. 
Therefore, our problem is reduced to determine the state $\ket{(\alpha) ; s}$.

In order to approach to the above-mentioned problem, we introduce the 
$su(3)$-algebra, the generators of which can be expressed in the form 
\begin{subequations}\label{2-8}
\begin{eqnarray}
& &{\hat R}_+(2)={\hat a}_2^*{\hat a}_3+{\hat b}_2^*{\hat b}_3 \ , \qquad
{\hat R}_+(1)={\hat a}_1^*{\hat a}_3+{\hat b}_1^*{\hat b}_3 \ , \nonumber\\
& &{\hat R}_-(2)={\hat a}_3^*{\hat a}_2+{\hat b}_3^*{\hat b}_2 \ , \qquad
{\hat R}_-(1)={\hat a}_3^*{\hat a}_1+{\hat b}_3^*{\hat b}_1 \ , 
\label{2-8a}\\
& &{\hat R}_2^1={\hat a}_1^*{\hat a}_2+{\hat b}_1^*{\hat b}_2 \ , \qquad
{\hat R}_1^2={\hat a}_2^*{\hat a}_1+{\hat b}_2^*{\hat b}_1 \ , 
\label{2-8b}\\
& &{\hat R}_2^2=(1/2)\cdot({\hat a}_2^*{\hat a}_2-{\hat a}_3^*{\hat a}_3)
+(1/2)\cdot({\hat b}_2^*{\hat b}_2-{\hat b}_3^*{\hat b}_3) \ , 
\nonumber\\
& &{\hat R}_1^1=(1/2)\cdot({\hat a}_1^*{\hat a}_1-{\hat a}_3^*{\hat a}_3)
+(1/2)\cdot({\hat b}_1^*{\hat b}_1-{\hat b}_3^*{\hat b}_3) \ . 
\label{2-8c}
\end{eqnarray}
\end{subequations}
Further, we introduce the following operators : 
\begin{subequations}\label{2-9}
\begin{eqnarray}
& &{\hat T}_+(2)={\hat a}_2^*{\hat b}_3^*-{\hat b}_2^*{\hat a}_3^* \ , \qquad
{\hat T}_+(1)={\hat a}_1^*{\hat b}_3^*-{\hat b}_1^*{\hat a}_3^* \ , \nonumber\\
& &{\hat T}_-(2)={\hat b}_3{\hat a}_2-{\hat a}_3{\hat b}_2 \ , \qquad
{\hat T}_-(1)={\hat b}_3{\hat a}_1-{\hat a}_3{\hat b}_1 \ , 
\label{2-9a}\\
& &{\hat T}_2^1={\hat a}_1^*{\hat a}_2+{\hat b}_1^*{\hat b}_2 \ , \qquad
{\hat T}_1^2={\hat a}_2^*{\hat a}_1+{\hat b}_2^*{\hat b}_1 \ , 
\label{2-9b}\\
& &{\hat T}_2^2=(1/2)\cdot({\hat a}_2^*{\hat a}_2+{\hat a}_3^*{\hat a}_3)
+(1/2)\cdot({\hat b}_2^*{\hat b}_2+{\hat b}_3^*{\hat b}_3)+1 \ , 
\nonumber\\
& &{\hat T}_1^1=(1/2)\cdot({\hat a}_1^*{\hat a}_1+{\hat a}_3^*{\hat a}_3)
+(1/2)\cdot({\hat b}_1^*{\hat b}_1+{\hat b}_3^*{\hat b}_3)+1 \ . 
\label{2-9c}
\end{eqnarray}
\end{subequations}
The set defined in the relation (\ref{2-9}) forms the $su(2,1)$-algebra. 
The above two algebras were already presented in Ref. \citen{8}.

We can show that in the present system, there exist various relations among 
the generators of the $su(2)$-, the $su(3)$- and the $su(2,1)$-algebras 
given in the above. Among them, the following relations are in particular 
useful for the later treatment : 
\begin{subequations}\label{2-10}
\begin{eqnarray}
& &[\ \hbox{\rm any\ of\ the\ $su(3)$-generators\ ,\ any\ of\ the\ 
$su(2)$-generators}\ ]=0 \ , \qquad
\label{2-10a}\\
& &[\ \hbox{\rm any\ of\ the\ $su(2,1)$-generators\ ,\ any\ of\ the\ 
$su(2)$-generators}\ ]=0 \ , \quad
\label{2-10b}
\end{eqnarray}
\end{subequations}
\begin{subequations}\label{2-11}
\begin{eqnarray}
& &[ {\hat R}_-(2) , {\hat T}_+(2)]=[{\hat R}_-(2) , {\hat T}_+(1) ]=0 \ , 
\nonumber\\
& &[ {\hat R}_-(1) , {\hat T}_+(2)]=[{\hat R}_-(1) , {\hat T}_+(1) ]=0 \ , 
\label{2-11a}\\
& &[ {\hat R}_+(2) , {\hat T}_+(2)]=[{\hat R}_+(1) , {\hat T}_+(1) ]=0 \ , 
\nonumber\\
& &[ {\hat R}_+(2) , {\hat T}_+(1)]=-[{\hat R}_+(1) , {\hat T}_+(2) ]
={\hat Q}_+ \neq 0 \ . 
\label{2-11b}
\end{eqnarray}
\end{subequations}
Here, ${\hat Q}_+$ is given as a component of a set of the operators 
$({\hat Q}_{\pm,0})$ defined by 
\begin{eqnarray}\label{2-12}
{\hat Q}_+&=&{\hat a}_1^*{\hat b}_2^*-{\hat b}_1^*{\hat a}_2^* \ , \qquad
{\hat Q}_- = {\hat b}_2{\hat a}_1-{\hat a}_2{\hat b}_1 \ , 
\nonumber\\
{\hat Q}_0&=&(1/2)\cdot({\hat a}_1^*{\hat a}_1+{\hat a}_2^*{\hat a}_2) 
+(1/2)\cdot({\hat b}_1^*{\hat b}_1+{\hat b}_2^*{\hat b}_2)+1 \nonumber\\
&=&(1/2)\cdot({\hat R}_1^1+{\hat R}_2^2)+(1/2)\cdot({\hat T}_1^1+{\hat T}_2^2)
\ . 
\end{eqnarray}
The set $({\hat Q}_{\pm,0})$ obeys the $su(1,1)$-algebra : 
\begin{equation}\label{2-13}
[{\hat Q}_+ , {\hat Q}_- ] = -2{\hat Q}_0 \ , \qquad
[{\hat Q}_0 , {\hat Q}_{\pm} ]=\pm {\hat Q}_\pm \ .
\end{equation}
From the above argument, we know that many-body systems consisting of six 
kinds of bosons are characterized not only by the $su(2)$-, the $su(3)$- 
and the $su(2,1)$-algebras but also by the $su(1,1)$-algebra. The operators 
${\hat Q}_\pm$ play a central role in \S\S 4 and 5. 
The above is the basic framework of our formalism.

\section{Determination of the intrinsic state}

We are now in a stage to determine the intrinsic state $\ket{(\alpha) ; s}$ 
obeying the condition (\ref{2-7}). As a possible candidate, we set up 
following form : 
\begin{subequations}\label{3-1}
\begin{eqnarray}
\ket{(\alpha) ; s}&=&
\left(\sqrt{p_1!p_2!}\right)^{-1}({\hat R}_+(1))^{p_1}({\hat R}_+(2))^{p_2}
\ket{(\beta) ; s} \ , 
\label{3-1x}\\
(\alpha)&=&p_1\ ,\ p_2\ , \ (\beta) \ ,\qquad
p_1,\ p_2=0,1,2,\cdots \ .
\label{3-1a}
\end{eqnarray}
\end{subequations}
Here, $(\beta)$ denotes a certain set of two numbers, which plays 
the same role as that of $(\alpha)$. There exists the relation (\ref{2-10a}) 
and $\ket{(\alpha) ; s}$ obeys the condition (\ref{2-7}). 
Therefore, the state $\ket{(\beta) ; s}$ should satisfy 
\begin{equation}\label{3-2}
{\hat S}_-\ket{(\beta) ; s}=0 \ , \qquad
{\hat S}_0\ket{(\beta) ; s}=-s\ket{(\beta) ; s} \ .
\end{equation}
Further, the form (\ref{3-1}) suggests us that ${\hat R}_+(1)$ and 
${\hat R}_+(2)$ plays a role of the raising operators from 
$\ket{(\beta) ; s}$ and, then, we set up the condition 
\begin{equation}\label{3-3}
{\hat R}_-(1)\ket{(\beta) ; s}=0 \ , \qquad
{\hat R}_-(2)\ket{(\beta) ; s}=0 \ .
\end{equation}
The state $\ket{(\beta);s}$ obeying the conditions (\ref{3-2}) and (\ref{3-3}) 
can be given as 
\begin {subequations}\label{3-4}
\begin{eqnarray}
\ket{(\beta);s}&=&\left(\sqrt{q_1!q_2!}\right)^{-1}
({\hat T}_+(1))^{q_1}({\hat T}_+(2))^{q_2}\ket{s} \ , 
\label{3-4x}\\
(\beta)&=&q_1 , \ q_2 , \qquad q_1, q_2=0,1,2,\cdots , 
\label{3-4a}
\end{eqnarray}
\end{subequations}
\vspace{-0.8cm}
\begin{eqnarray}\label{3-5}
\ket{s}&=&\left(\sqrt{(2s)!}\right)^{-1}({\hat b}_3^*)^{2s}\ket{0} \ . 
\end{eqnarray}
Therefore, combining with the form (\ref{3-1}), we have 
\begin{eqnarray}\label{3-6}
\ket{p_1p_2q_1q_2 ; s}
&=&\left(\sqrt{p_1!p_2!}\right)^{-1}({\hat R}_+(1))^{p_1}({\hat R}_+(2))^{p_2}
\nonumber\\
& &\times \left(\sqrt{q_1!q_2!}\right)^{-1}({\hat T}_+(1))^{q_1}
({\hat T}_+(2))^{q_2}\cdot \left(\sqrt{(2s)!}\right)^{-1}
({\hat b}_3^*)^{2s}\ket{0} \ .\qquad
\end{eqnarray}
The state (\ref{3-6}) can be regarded as the intrinsic state and forms the 
linearly independent set, but, it is not orthogonal. 
The numbers $(p_1,p_2,q_1,q_2)$ do not play a role of the quantum numbers. 
Then, by making an appropriate linear combination for the 
states (\ref{3-6}), we can construct the orthogonal set.

For the above task, we must prepare four mutually commuted hermitian 
operators expressed in terms of the $su(3)$-generators (\ref{2-8a})$\sim$
(\ref{2-8c}). Then, including ${\hat {\mib S}}^2$ and ${\hat S}_0$, we 
have six mutually commuted hermitian operators. 
First, we define the operators 
\begin{eqnarray}\label{3-7}
{\hat M}_+&=&{\hat a}_1^*{\hat a}_2+{\hat b}_1^*{\hat b}_2
={\hat R}_2^1={\hat T}_2^1 \ , \nonumber\\
{\hat M}_-&=&{\hat a}_2^*{\hat a}_1+{\hat b}_2^*{\hat b}_1
={\hat R}_1^2={\hat T}_1^2 \ , \nonumber\\
{\hat M}_0&=&(1/2)\cdot({\hat a}_1^*{\hat a}_1-{\hat a}_2^*{\hat a}_2)
+(1/2)\cdot({\hat b}_1^*{\hat b}_1-{\hat b}_2^*{\hat b}_2)\nonumber\\
&=&{\hat R}_1^1-{\hat R}_2^2={\hat T}_1^1-{\hat T}_2^2 \ .
\end{eqnarray}
As can be seen from the form, the set $({\hat M}_{\pm,0})$ obeys 
the $su(2)$-algebra and, then, as mutually commuted operators, 
we can choose ${\hat M}_0$ and the Casimir operator ${\hat {\mib M}}^2$ 
expressed in the form 
\begin{equation}\label{3-8}
{\hat {\mib M}}^2={\hat M}_0^2+(1/2)\cdot({\hat M}_+{\hat M}_-
+{\hat M}_-{\hat M}_+) \ .
\end{equation}
Next, we define the operator 
\begin{eqnarray}\label{3-9}
{\hat R}_0&=&(1/2)\cdot({\hat a}_1^*{\hat a}_1+{\hat a}_2^*{\hat a}_2)
+(1/2)\cdot({\hat b}_1^*{\hat b}_1+{\hat b}_2^*{\hat b}_2)
-({\hat a}_3^*{\hat a}_3+{\hat b}_3^*{\hat b}_3) \nonumber\\
&=&{\hat R}_1^1+{\hat R}_2^2
={\hat S}(1)+{\hat S}(2)-2{\hat S}(3) \ . 
\end{eqnarray}
Finally, the Casimir operator for the $su(3)$-algebra, which we denote as 
${\hat {\mib R}}^2$, is introduced : 
\begin{eqnarray}\label{3-10x}
{\hat {\mib R}}^2&=&
(4/3)\cdot[({\hat R}_1^1)^2-{\hat R}_1^1{\hat R}_2^2+({\hat R}_2^2)^2]
+(1/2)\cdot ({\hat R}_2^1{\hat R}_1^2+{\hat R}_1^2{\hat R}_2^1) \nonumber\\
& &+(1/2)\cdot[{\hat R}_+(1){\hat R}_-(1)+{\hat R}_+(2){\hat R}_-(2)
+{\hat R}_-(1){\hat R}_+(1)+{\hat R}_-(2){\hat R}_+(2)] \nonumber\\
&=&(1/3)\cdot{\hat R}_0({\hat R}_0-3)+{\hat {\mib M}}^2+
({\hat R}_+(1){\hat R}_-(1)+{\hat R}_+(2){\hat R}_-(2)) \ .
\end{eqnarray}
The above operators ${\hat {\mib M}}^2$, ${\hat M}_0$, ${\hat {\mib R}}^2$ 
and ${\hat R}_0$ are the mutually commuted hermitian operators and, 
of course, they commute with ${\hat {\mib S}}^2$ and ${\hat S}_0$.

The operators $({\hat R}_+(1), {\hat R}_+(2))$ and 
$({\hat T}_+(1) , {\hat T}_+(2))$ obey the relations 
\begin{subequations}\label{3-10}
\begin{eqnarray}
& &[{\hat M}_+ , {\hat R}_+(1)]=0 \ , \qquad
[{\hat M}_+ , {\hat R}_+(2)]={\hat R}_+(1) \ , \nonumber\\
& &[{\hat M}_- , {\hat R}_+(1)]={\hat R}_+(2) \ , \qquad
[{\hat M}_- , {\hat R}_+(2)]=0 \ , \nonumber\\
& &[{\hat M}_0 , {\hat R}_+(1)]=+1/2\cdot{\hat R}_+(1) \ , \qquad
[{\hat M}_0 , {\hat R}_+(2)]=-1/2\cdot {\hat R}_+(2) \ , 
\label{3-10a}\\
& &[{\hat M}_+ , {\hat T}_+(1)]=0 \ , \qquad
[{\hat M}_+ , {\hat T}_+(2)]={\hat T}_+(1) \ , \nonumber\\
& &[{\hat M}_- , {\hat T}_+(1)]={\hat T}_+(2) \ , \qquad
[{\hat M}_- , {\hat T}_+(2)]=0 \ , \nonumber\\
& &[{\hat M}_0 , {\hat T}_+(1)]=+1/2\cdot{\hat T}_+(1) \ , \qquad
[{\hat M}_0 , {\hat T}_+(2)]=-1/2\cdot {\hat T}_+(2) \ , 
\label{3-10b}
\end{eqnarray}
\end{subequations}
The relations (\ref{3-10a}) and (\ref{3-10b}) tell us that the operators 
$({\hat R}_+(1),{\hat R}_+(2))$ and $({\hat T}_+(1),{\hat T}_+(2))$ 
denote the spherical tensors with rank $j=1/2$, i.e., 
$({\hat R}_+(1)={\hat R}_{1/2,1/2}$, ${\hat R}_+(2)={\hat R}_{1/2,-1/2})$ and 
$({\hat T}_+(1)={\hat T}_{1/2,1/2}$, ${\hat T}_+(2)={\hat T}_{1/2,-1/2})$. 
Therefore, we can construct the spherical tensors with rank $k$ and the 
$k_0$-th component and with rank $l$ and the $l_0$-th component, respectively, 
in the following form : 
\begin{subequations}\label{3-11}
\begin{eqnarray}
& &{\hat R}_+(k,k_0)=\left(\sqrt{(k+k_0)!(k-k_0)!}\right)^{-1}
\cdot ({\hat R}_+(1))^{k+k_0}({\hat R}_+(2))^{k-k_0} \ , 
\label{3-11a}\\
& &{\hat T}_+(l,l_0)=\left(\sqrt{(l+l_0)!(l-l_0)!}\right)^{-1}
\cdot ({\hat T}_+(1))^{l+l_0}({\hat T}_+(2))^{l-l_0} \ , 
\label{3-11b}
\end{eqnarray}
\end{subequations}
\vspace{-0.8cm}
\begin{subequations}\label{3-12}
\begin{eqnarray}
& &k=0, 1/2, 1, 3/2, \cdots , \qquad
k_0=-k, -k+1, \cdots , k-1, k \ , 
\label{3-12a}\\
& &l=0, 1/2, 1, 3/2, \cdots , \qquad
l_0=-l, -l+1, \cdots , l-1, l \ , 
\label{3-12b}
\end{eqnarray}
\end{subequations}
The boson operators ${\hat a}_3^*$ and ${\hat b}_3^*$ are scalars with respect 
to the $su(2)$-algebra $({\hat M}_{\pm, 0})$. The operator ${\hat R}_0$ 
obeys the relations 
\begin{subequations}\label{3-13}
\begin{eqnarray}
& &[{\hat R}_0 , {\hat R}_+(1) ]=+3/2\cdot{\hat R}_+(1) \ , \qquad
[{\hat R}_0 , {\hat R}_+(2) ]=+3/2\cdot{\hat R}_+(2) \ , 
\label{3-13a}\\
& &[{\hat R}_0 , {\hat T}_+(1) ]=-1/2\cdot{\hat T}_+(1) \ , \qquad
[{\hat R}_0 , {\hat T}_+(2) ]=-1/2\cdot{\hat T}_+(2) \ . 
\label{3-13b}
\end{eqnarray}
\end{subequations}
Therefore, we have 
\begin{subequations}\label{3-14}
\begin{eqnarray}
& &[{\hat R}_0 , {\hat R}_+(k,k_0) ]=+3k\cdot{\hat R}_+(k,k_0) \ , 
\label{3-14a}\\
& &[{\hat R}_0 , {\hat T}_+(l,l_0) ]=-l\cdot{\hat T}_+(l,l_0) \ .
\label{3-14b}
\end{eqnarray}
\end{subequations}
Further, there exists the relation 
\begin{equation}\label{3-15}
[{\hat {\mib R}}^2 ,{\hat R}_+(k,k_0)]=0 \ .
\end{equation}

Under the above preparation, let us construct the orthogonal set for the 
intrinsic state. For the state (\ref{3-6}), we put 
\begin{equation}\label{3-16}
p_1=k+k_0 \ , \qquad p_2=k-k_0 \ , \qquad
q_1=l+l_0 \ , \qquad q_2=l-l_0 \ .
\end{equation}
By superposing the state (\ref{3-6}) under the Clebsh-Gordan(CG)-coefficients, 
we define the state as 
\begin{equation}\label{3-17}
\ket{k,l,m,m_0;s}
=\sum_{k_0,l_0}\bra{kk_0ll_0}mm_0\rangle{\hat R}_+(k,k_0){\hat T}_+(l,l_0)
\ket{s} \ .
\end{equation}
The state $\ket{k,l,m,m_0;s}$ satisfies the eigenvalue equations for the 
operators ${\hat {\mib R}}^2$, ${\hat R}_0$, ${\hat {\mib M}}^2$ and 
${\hat M}_0$, the eigenvalues of which are given as 
\begin{subequations}\label{3-18}
\begin{eqnarray}
(1/3)(2s+l)(2s+l+3)+l(l+1) \qquad &\hbox{\rm for}&\quad {\hat {\mib R}}^2 \ , 
\label{3-18a}\\
3k-(2s+l) \qquad &\hbox{\rm for}&\quad {\hat R}_0 \ , 
\label{3-18b}\\
m(m+1) \qquad &\hbox{\rm for}&\quad {\hat {\mib M}}^2 \ , 
\label{3-19a}\\
m_0 \qquad &\hbox{\rm for}&\quad {\hat M}_0 \ . 
\label{3-19b}
\end{eqnarray}
\end{subequations}
Under a given value of $s$, we can determine $(k,l,m,m_0)$. 
From the relation (\ref{3-18a}), $l$ is determined and, then, the relation 
(\ref{3-18b}) give us $k$. Further, the relations (\ref{3-19a}) and 
(\ref{3-19b}) give us $m$ and $m_0$. It may be self-evident that the 
set of the states $\ket{k,l,m,m_0;s}$ forms the orthogonal set. 

From the above argument, we have the state 
\begin{eqnarray}\label{3-20}
\ket{k,l,m,m_0;s,s_0}
&=&\left(\sqrt{\Gamma}\right)^{-1}\cdot({\hat S}_+)^{s+s_0}\cdot
\sum_{k_0,l_0}C(k,k_0;l,l_0|m,m_0) \nonumber\\
& &\times\left(\sqrt{(k+k_0)!(k-k_0)!}\right)^{-1}({\hat R}_+(1))^{k+k_0}
({\hat R}_+(2))^{k-k_0} \nonumber\\
& &\times\left(\sqrt{(l+l_0)!(l-l_0)!}\right)^{-1}({\hat T}_+(1))^{l+l_0}
({\hat T}_+(2))^{l-l_0} \nonumber\\
& &\times \left(\sqrt{(2s)!}\right)^{-1}({\hat b}_3^*)^{2s} \ket{0} \ .
\end{eqnarray}
Here, $\Gamma$ and $C(k,k_0;l,l_0|m,m_0)$ denote 
the normalization constant and the Clebsch-Gordan coefficient, respectively.

\section{Reformation of the orthogonal set}

Main task in this section is to show the eigenstate $\ket{k,l,m,m_0;s,s_0}$, 
the form of which is different from that given in the relation (\ref{3-20}), 
but essentially, the same. 
The form presented in this section may be  useful 
for constructing the coherent state of the present system. 
In the case $(s_0\!=\!-s,\ m_0\!=\!-m,\ m\!=\!k+l)$, 
the state $\ket{k,l,m,m_0;s,s_0}$ 
can be expressed as 
\begin{equation}\label{4-1}
\ket{k,l,(k+l),-(k+l); s,-s}
=({\hat R}_+(2))^{2k}({\hat T}_+(2))^{2l}({\hat b}_3^*)^{2s}\ket{0} 
\ (=\ket{2k,2l,2s}) \ .
\end{equation}
In the above expression, including the normalization constant $\Gamma$, 
the factorial factors are omitted. It may be noted that the case 
$(m_0=-m, m=k+l)$ corresponds to $(k_0=-k, l_0=-l)$ and 
$C(k,-k;l,-l|(k+l),-(k+l))=1$. 
The state (\ref{4-1}) obeys the conditions 
\begin{eqnarray}
& &{\hat S}_-\ket{2k,2l,2s}=0 \ , \nonumber\\
& &{\hat {\mib S}}^2\ket{2k,2l,2s}=s(s+1)\ket{2k,2l,2s} \ , \nonumber\\
& &{\hat S}_0\ket{2k,2l,2s}=-s\ket{2k,2l,2s} \ , 
\label{4-2}\\
& &{\hat M}_-\ket{2k,2l,2s}=0 \ , \nonumber\\
& &{\hat {\mib M}}^2\ket{2k,2l,2s}=(k+l)(k+l+1)\ket{2k,2l,2s} \ , \nonumber\\
& &{\hat M}_0\ket{2k,2l,2s}=-(k+l)\ket{2k,2l,2s} \ , 
\label{4-3}\\
& &{\hat R}_0\ket{2k,2l,2s}=(3k-(2s+l))\ket{2k,2l,2s} \ . 
\label{4-4}
\end{eqnarray}
The relation (\ref{4-3}) presents a key to arrive at our final goal. 
By shifting $2k$ and $2l$ to $2k-(k+l-m)\ (=k-l+m)$ and 
$2l-(k+l-m)\ (=l-k+m)$, respectively, we have 
\begin{equation}\label{4-5}
\ket{k-l+m, l-k+m, 2s}=
({\hat R}_+(2))^{k-l+m}({\hat T}_+(2))^{l-k+m}({\hat b}_3^*)^{2s}\ket{0} \ .
\end{equation}
Of course, the state (\ref{4-5}) satisfies the same relation as that given 
in Eq. (\ref{4-2}). Concerning the relations (\ref{4-3}) and (\ref{4-4}), 
we have 
\begin{eqnarray}
& &{\hat M}_-\ket{k\!-\!l\!+\!m, l\!-\!k\!+\!m, 2s}=0 \ , \nonumber\\
& &{\hat {\mib M}}^2\ket{k\!-\!l\!+\!m, l\!-\!k\!+\!m, 2s}
=m(m+1)\ket{k\!-\!l\!+\!m, l\!-\!k\!+\!m, 2s} 
\ , \nonumber\\
& &{\hat M}_0\ket{k\!-\!l\!+\!m, l\!-\!k\!+\!m, 2s}
=-m\ket{k\!-\!l\!+\!m, l\!-\!k\!+\!m, 2s} \ , 
\label{4-6}\\
& &{\hat R}_0\ket{k\!-\!l\!+\!m, l\!-\!k\!+\!m, 2s}=[3k-(2s+l)-(k+l-m)]
\ket{k\!-\!l\!+\!m, l\!-\!k\!+\!m, 2s} \ . \nonumber\\
& &\label{4-7}
\end{eqnarray}
The state $\ket{k-l+m,l-k+m,2s}$ is an eigenstate of ${\hat R}_0$ 
with the eigenvalue shown in the relation (\ref{4-7}). Then, let us 
show that the eigenstate of ${\hat R}_0$ with the eigenvalue $3k-(2s+l)$ 
is obtained by the operator ${\hat Q}_+$ defined in the relation 
(\ref{2-12}) : 
\begin{equation}\label{4-8}
{\hat Q}_+=[{\hat R}_+(2),{\hat T}_+(1)]
=-[{\hat R}_+(1) , {\hat T}_+(2)]
={\hat a}_1^*{\hat b}_2^*-{\hat b}_1^*{\hat a}_2^* \ .
\end{equation}
The operator ${\hat Q}_+$ obeys the relations 
\begin{eqnarray}
& &[{\hat M}_{\pm} , {\hat Q}_+ ]=[{\hat M}_0 , {\hat Q}_+]=0 \ , 
\label{4-9}\\
& &[{\hat R}_0 , {\hat Q}_+ ]={\hat Q}_+ \ . 
\label{4-10}
\end{eqnarray}
Then, we define the state 
\begin{eqnarray}\label{4-11}
\ket{k,l,m,s}&=&
({\hat Q}_+)^{k+l-m}\ket{k-l+m,l-k+m,2s} \nonumber\\
&=&({\hat Q}_+)^{k+l-m}({\hat R}_+(2))^{k-l+m}
({\hat T}_+(2))^{l-k+m}({\hat b}_3^*)^{2s} \ket{0} \ . 
\end{eqnarray}
The state (\ref{4-11}) obeys the relations 
\begin{eqnarray}
& &{\hat M}_-\ket{k, l, m, s}=0 \ , \nonumber\\
& &{\hat {\mib M}}^2\ket{k, l, m, s}=m(m+1) \ket{k, l, m, s}
\ , \nonumber\\
& &{\hat M}_0\ket{k, l, m, s}=-m\ket{k, l, m, s} \ , 
\label{4-12}\\
& &{\hat R}_0\ket{k, l, m, s}=[3k-(2s+l)]\ket{k, l, m, s} \ . 
\label{4-13}
\end{eqnarray}
The operators ${\hat Q}_+$, ${\hat R}_+(2)$ and ${\hat T}_+(2)$ are 
mutually commuted and, then, the ordering of these operators 
appearing in the state (\ref{4-11}) is arbitrary. The Casimir operator 
${\hat {\mib R}}^2$ given in the relation (\ref{3-10x}) satisfies 
\begin{subequations}\label{4-14}
\begin{eqnarray}
& &{\hat {\mib R}}^2\ket{k,l,m,s}
=({\hat R}_+(2))^{k-l+m}\cdot {\hat {\mib R}}^2 \kket{k,l,m,s} \ , 
\label{4-14x}\\
& &\kket{k,l,m,s}=({\hat T}_+(2))^{l-k+m}({\hat Q}_+)^{k+l-m}
({\hat b}_3^*)^{2s}\ket{0} \ . 
\label{4-14a}
\end{eqnarray}
\end{subequations}
The state $\kket{k,l,m,s}$ satisfies 
\begin{subequations}\label{4-15}
\begin{eqnarray}
& &{\hat R}_0\kket{k,l,m,s}=[(l+3(k-m))/2-2s]\kket{k,l,m,s} \ , 
\label{4-15a}\\
& &{\hat {\mib M}}^2\kket{k,l,m,s}
=[(l-k+m)/2][(l-k+m)/2+1]\kket{k,l,m,s} \ , 
\label{4-15b}\\
& &[{\hat R}_+(1){\hat R}_-(1)\!+\!{\hat R}_+(2){\hat R}_-(2)]\kket{k,l,m,s}
=(k\!+\!l\!-\!m)(l\!-\!k\!+\!m\!+\!2s\!+\!2)\kket{k,l,m,s} \ . 
\nonumber\\
& &\label{4-15c}
\end{eqnarray}
\end{subequations}
The relations (\ref{4-15a})$\sim$(\ref{4-15c}) lead us to the following 
result : 
\begin{equation}\label{4-16}
{\hat {\mib R}}^2\kket{k,l,m,s}
=[(1/3)\cdot (2s+l)(2s+l+3)+l(l+1)]\kket{k,l,m,s} \ .
\end{equation}
As can be seen in the above relation, $\lambda$ and $\mu$ appearing in the 
Elliott's $su(3)$-model correspond to $\lambda=2s+l$ and $\mu=l$.\cite{14} 
From the above argument, the state defined in the following can be 
regarded as the intrinsic state introduced in \S 3 : 
\begin{equation}\label{4-17}
\ket{k,l,m,m_0,s}=({\hat M}_+)^{m+m_0}\ket{k,l,m,s} \ .
\end{equation}
Thus, we are able to have the state $\ket{k,l,m,m_0;s,s_0}$ in the form 
\begin{eqnarray}\label{4-18}
\ket{k,l,m,m_0;s,s_0}&=&
\left(\sqrt{\Gamma}\right)^{-1}({\hat S}_+)^{s+s_0}({\hat M}_+)^{m+m_0}
\nonumber\\
& &\times ({\hat Q}_+)^{k+l-m}({\hat R}_+(2))^{k-l+m}
({\hat T}_+(2))^{l-k+m}({\hat b}_3^*)^{2s} \ket{0} \ .\qquad
\end{eqnarray}
In contrast to the form (\ref{3-20}), the form (\ref{4-18}) is a monomial. 
We can see that the state (\ref{4-18}) is constructed in terms of the 
operators ${\hat S}_+$, ${\hat M}_+$, ${\hat Q}_+$, ${\hat T}_+(2)$ 
and ${\hat R}_+(2)$. In this sense, these operators may be called the 
state generating operators. The quantity $\Gamma$ denotes the normalization 
constant which may be different from $\Gamma$ in the state (\ref{3-20}). 
Straightforward calculation gives us $\Gamma$ in the following form : 
\begin{eqnarray}\label{4-19}
\Gamma&=&[(2s)!(s+s_0)!/(s-s_0)!][(2m)!(m+m_0)!/(m-m_0)!] \nonumber\\
& &\times [(2s)!(k+l-m)!(l-k+m)!/(2s-k+l-m)!]^2 \nonumber\\
& &\times \sum_{i=0}^{k+l-m}\sum_{j=0}^{l-k+m}
[(2k-i+j)!(l-k+m+i-j)!(2(s+l-k)-j)!] \nonumber\\
& &\qquad\qquad\qquad\qquad 
\times [i!j!(k+l-m-i)!((l-k+m-j)!)^2]^{-1} \ .
\end{eqnarray}

Until the present, we did not contact with the conditions which the 
quantum numbers $(k,l,m,m_0,s,s_0)$ should satisfy. These numbers cannot 
change in arbitrary regions. Finally, we discuss this problem. 
For this aim, with the use of the definition of ${\hat R}_+(2)$ shown in 
the relation (\ref{2-8a}), we rewrite the state (\ref{4-18}) in the form 
\begin{eqnarray}\label{4-20}
\ket{k,l,m,m_0;s,s_0}&=&
\left(\sqrt{\Gamma}\right)^{-1}(2s)!/(2s-k+l-m)! \nonumber\\
& &\times ({\hat S}_+)^{s+s_0}({\hat M}_+)^{m+m_0}
({\hat Q}_+)^{k+l-m}({\hat T}_+(2))^{l-k+m} \nonumber\\
& &\times ({\hat b}_2^*)^{k-l+m}({\hat b}_3^*)^{2s-k+l-m} \ket{0} \ .
\end{eqnarray}
We can see that the state $\ket{k,l,m,m_0;s,s_0}$ can be defined in the 
regions 
\begin{subequations}\label{4-21}
\begin{eqnarray}\label{4-21x}
& &-s \le s_0 \le +s \ , \qquad -m \le m_0 \le +m \ , 
\nonumber\\
& &k+l-m \ge 0 \ , \quad 
l-k+m \ge 0 \ , \quad k-l+m \ge 0 \ , \quad 2s-k+l-m \ge 0 \ , \nonumber\\
& &
\end{eqnarray}
that is, 
\begin{equation}
|s_0| \le s \ , \quad |m_0| \le m \ , \quad 
|k-l|\le m \le k+l \ , \quad 
m \le 2s+l-k \ .
\label{4-21a}
\end{equation}
\end{subequations}
The relation (\ref{4-21}) is discussed again in \S 5.

\section{Discussion}

It may be interesting to investigate the properties of the orthogonal 
set\break 
$\{ \ket{k,l,m,m_0;s,s_0} \}$ in the language of the conventional 
manner for three $su(2)$-spins, i.e., the successive addition of these spins. 
If we follow the conventional manner, the orthogonal set is specified 
in terms of the quantum numbers $(s_1,s_2,s_{12},s_3 ; s,s_0)$. 
Here, $s_i$ ($i=1,2,3)$ denotes the magnitude of the $i$-th $su(2)$-spin 
and $s_{12}$ the magnitude of the $su(2)$-spin coupled by the first 
and the second $su(2)$-spin. The operator expressing the magnitude of 
the $i$-th $su(2)$-spin can be found in the relation (\ref{2-2b}). 
The operation of ${\hat S}(i)$ on the state $\ket{k,l,m,m_0;s,s_0}$ 
leads us to 
%\begin{subequations}\label{5-1}
\begin{equation}\label{5-1x}
{\hat S}(i)\ket{k,l,m,m_0;s,s_0}=s_i\ket{k,l,m,m_0;s,s_0}\ . \quad
(i=1,2,3)
\end{equation}
Here, $s_i$ is obtained in the form 
\setcounter{equation}{0}
\begin{subequations}\label{5-1}
\begin{eqnarray}
& &s_1=(1/2)\cdot (k+l+m_0) \ , 
\label{5-1a}\\
& &s_2=(1/2)\cdot (k+l-m_0) \ , 
\label{5-1b}\\
& &s_3=l-k+s \ . 
\label{5-1c}
\end{eqnarray}
\end{subequations}
In the case of the quantum number $s_{12}$, we must treat the Casimir 
operator of the $su(2)$-spin coupled by the first and the second 
$su(2)$-spin, which we denote as ${\hat {\mib S}}_{12}^2$ : 
\begin{eqnarray}\label{5-2}
{\hat {\mib S}}_{12}^2&=&
({\hat S}_0(1)+{\hat S}_0(2))^2 
+(1/2)\cdot [({\hat S}_+(1)+{\hat S}_+(2))
({\hat S}_-(1)+{\hat S}_-(2)) \nonumber\\
& &\qquad\qquad\qquad\qquad\qquad\qquad
+({\hat S}_-(1)+{\hat S}_-(2))({\hat S}_+(1)
+{\hat S}_+(2))] \ . \qquad
\end{eqnarray}
Substituting the definition (\ref{2-2a}) into the form (\ref{5-2}), 
we have 
\begin{equation}\label{5-3}
{\hat {\mib S}}_{12}^2
=({\hat S}(1)+{\hat S}(2))({\hat S}(1)+{\hat S}(2)+1)-{\hat Q}_+{\hat Q}_- \ .
\end{equation}
On the other hand, with use of the definition (\ref{3-7}), we obtain 
the relation 
\begin{equation}\label{5-4}
{\hat {\mib M}}^2={\hat {\mib S}}_{12}^2 \ . 
\end{equation}
Therefore, $s_{12}$ is equal to $m$ :
\begin{equation}\label{5-1d}
s_{12}=m \ .
\end{equation}
Inversely, the relations (\ref{5-1a})$\sim$(\ref{5-1c}) and (\ref{5-1d}) 
give us the relations 
\begin{subequations}\label{5-5}
\begin{eqnarray}
& &k=(1/2)\cdot (s_1+s_2-s_3+s) \ , 
\label{5-5a}\\
& &l=(1/2)\cdot (s_1+s_2+s_3-s) \ , 
\label{5-5b}\\
& &m=s_{12} \ , 
\label{5-5c}\\
& &m_0=s_1-s_2 \ .
\label{5-5d}
\end{eqnarray}
\end{subequations}
Substituting the relation (\ref{5-5}) into the inequality (\ref{4-21}), 
we get the following inequality : 
\begin{equation}\label{5-6}
|s_0|\le s \ , \quad 
|s_1-s_2|\le s_{12}\le s_1+s_2 \ , \quad 
|s_{12}-s_3|\le s\le s_{12}+s_3 \ . 
\end{equation}
The above is nothing but the coupling rule for three $su(2)$-spins. 
Thus, we can summarize the state $\ket{s_1,s_2,(s_{12}),s_3;s,s_0}$ 
as follows : 
\begin{eqnarray}
& &\ket{s_1,s_2,(s_{12}),s_3;s,s_0}
=\left(\sqrt{\Gamma}\right)^{-1}({\hat S}_+)^{s+s_0}
({\hat M}_+)^{s_1-s_2+s_{12}}({\hat Q}_+)^{s_1+s_2-s_{12}} \nonumber\\
& &\qquad\qquad\qquad\qquad\qquad
\times ({\hat T}_+(2))^{s_{12}+s_3-s}({\hat R}_+(2))^{s_{12}-s_3+s}
({\hat b}_3^*)^{2s} \ket{0} \ . 
\label{5-7}\\
& &\Gamma=
[(2s)!(s+s_0)!/(s-s_0)!][(2s_{12})!(s_1-s_2+s_{12})!/(s_2-s_1+s_{12})!]
\nonumber\\
& &\qquad
\times [(2s)!(s_1+s_2-s_{12})!(s_{12}+s_3-s)!/(s_3-s_{12}+s)!]^2 
\nonumber\\
& &\qquad
\times \sum_{i=0}^{s_1+s_2-s_{12}}\sum_{j=0}^{s_{12}+s_3-s}
[(s_1+s_2-s_3+s-i+j)!(s_{12}+s_3-s+i-j)!(2s_3-j)!] \nonumber\\
& &\qquad\qquad\qquad\qquad\qquad \times
[i!j!(s_1+s_2-s_{12}-i)!((s_{12}+s_3-s-j)!)^2]^{-1} \ .
\end{eqnarray}

In this paper, we gave the orthogonal set for describing the systems 
consisting of six kinds of boson operators, mainly, in relation to 
the $su(2)$- and the $su(3)$-algebras. We show that the orthogonal set 
obtained in this paper can be applied to the case of the $su(2,1)$-algebra. 
In addition to the operators ${\hat {\mib S}}^2$ and ${\hat S}_0$, 
four mutually commuted hermitian operators must be introduced. 
As was shown in the relation (\ref{3-7}), the operators $({\hat M}_{\pm,0})$ 
can be also expressed in terms of the $su(2,1)$-generators : 
\begin{equation}\label{5-9}
{\hat M}_+={\hat T}_2^1 \ , \qquad {\hat M}_-={\hat T}_1^2 \ , \qquad 
{\hat M}_0={\hat T}_1^1-{\hat T}_2^2 \ .
\end{equation}
Then, ${\hat {\mib M}}^2$ and ${\hat M}_0$ are also useful in the 
present case. Instead of ${\hat R}_0$, we introduce the operator ${\hat T}_0$ 
in the form 
\begin{eqnarray}\label{5-10}
{\hat T}_0={\hat T}_1^1+{\hat T}_2^2
&=& (1/2)\cdot({\hat a}_1^*{\hat a}_1+{\hat a}_2^*{\hat a}_2) \nonumber\\
& &+(1/2)\cdot({\hat b}_1^*{\hat b}_1+{\hat b}_2^*{\hat b}_2) + 
({\hat a}_3^*{\hat a}_3+{\hat b}_3^*{\hat b}_3)+2 \nonumber\\
&=&{\hat S}(1)+{\hat S}(2)+2{\hat S}(3)+2 \ .  
\end{eqnarray}
Further, the Casimir operator of the $su(2,1)$-algebra, which we denote 
as ${\hat {\mib T}}^2$, is useful : 
\begin{eqnarray}\label{5-11}
{\hat {\mib T}}^2&=&
(4/3)\cdot[({\hat T}_1^1)^2-{\hat T}_1^1{\hat T}_2^2+({\hat T}_2^2)^2]
+(1/2)\cdot ({\hat T}_2^1{\hat T}_1^2+{\hat T}_1^2{\hat T}_2^1) \nonumber\\
& &-(1/2)\cdot[{\hat T}_+(1){\hat T}_-(1)+{\hat T}_+(2){\hat T}_-(2)
+{\hat T}_-(1){\hat T}_+(1)+{\hat T}_-(2){\hat T}_+(2)]  \ . \qquad
\end{eqnarray}
The above operator can be rewritten as 
\begin{eqnarray}\label{5-12}
{\hat {\mib T}}^2
&=&{\hat {\mib R}}^2-(1/6)\cdot({\hat T}_0^2-{\hat R}_0^2) \nonumber\\
&=&{\hat {\mib R}}^2-(2/3)\cdot ({\hat S}(1)+{\hat S}(2)+1)
(2{\hat S}(3)+1) \ . \qquad
\end{eqnarray}
From the above argument, we can see that the state (\ref{4-18}) or (\ref{5-7}) 
is the eigenstate of ${\hat {\mib M}}^2$, ${\hat M}_0$, ${\hat {\mib T}}^2$ 
and ${\hat T}_0$. The eigenvalues of ${\hat {\mib M}}^2$ and ${\hat M}_0$ 
are naturally $m(m+1)$ and $m_0$, respectively. 
The cases of the eigenvalues of ${\hat {\mib T}}^2$ and ${\hat T}_0$, 
which we denote as $({\mib T}^2)_{\rm ev}$ and $(T_0)_{\rm ev}$, 
are calculated in the 
following form : 
\begin{eqnarray}
({\mib T}^2)_{\rm ev}
&=& 
(1/3)\cdot(2s-k+2)(2s-k-1)+k(k+1) \nonumber\\
&=&(1/3)\cdot(-2(s+1)+k)(-2(s+1)+k+3)+k(k+1) \ , 
%\nonumber\\
%& &
\label{5-13}\\
(T_0)_{\rm ev}
&=& 
3l+2s-k+2 \nonumber\\
&=& 3l-(-2(s+1)+k) \ . 
\label{5-14}
\end{eqnarray}
It is interesting to see that if $s$, $k$ and $l$ in the expressions for 
the eigenvalues of ${\hat {\mib R}}^2$ and ${\hat R}_0$ shown in the 
relation (\ref{3-18}) are replaced with 
$-(s+1)$, $l$ and $k$, respectively, the results are reduced to the relations 
(\ref{5-13}) and (\ref{5-14}). From the above argument, we can understand 
that the orthogonal set obtained in this paper is also applicable 
to the case of the $su(2,1)$-algebra.

Until the present stage, we have discussed three interesting aspects of the 
orthogonal set (\ref{3-20}) 
in relation to the $su(3)$-, the $su(2)\times su(2)\times su(2)$ and 
the $su(2,1)$-algebra. 
We, further, show fourth aspect related with 
the role of the operators $({\hat Q}_{\pm,0})$. 
As was shown in the relation (\ref{2-13}), the set $({\hat Q}_{\pm,0})$ 
obeys the $su(1,1)$-algebra and, then, the Casimir operator 
${\hat {\mib Q}}^2$ is defined in the form 
\begin{equation}\label{5-15}
{\hat {\mib Q}}^2={\hat Q}_0^2-(1/2)\cdot
({\hat Q}_+{\hat Q}_-+{\hat Q}_-{\hat Q}_+)
={\hat Q}_0({\hat Q}_0-1)-{\hat Q}_+{\hat Q}_- \ .
\end{equation}
Since ${\hat Q}_0={\hat S}(1)+{\hat S}(2)+1$, the operator 
${\hat {\mib Q}}^2$ is reduced to ${\hat {\mib S}}_{12}^2$ shown in the 
relation (\ref{5-3}). Then, the relation (\ref{5-4}) gives us 
\begin{equation}\label{5-16}
{\hat {\mib Q}}^2={\hat {\mib M}}^2 \ . 
\end{equation}
Further, we introduce the operator ${\hat N}_0$ defined as 
\begin{eqnarray}\label{5-17}
{\hat N}_0&=&(1/2)\cdot({\hat a}_3^*{\hat a}_3+{\hat b}_3^*{\hat b}_3) 
\nonumber\\
&=&(1/4)\cdot (({\hat T}_1^1+{\hat T}_2^2)-({\hat R}_1^1+{\hat R}_2^2))-1/2 \ .
\end{eqnarray}
In this way, we can prepare new set of six mutually commuted 
hermitian operators\break 
$({\hat {\mib S}}^2, {\hat S}_0 , {\hat {\mib Q}}^2, {\hat Q}_0 , 
{\hat M}_0 , {\hat N}_0)$. The eigenvalue problem for these operators can be 
easily solved. First, we note the following relation : 
\begin{eqnarray}\label{5-18}
& &{\hat Q}_-\ket{k-l+m,l-k+m,2s}=0 \ , \nonumber\\
& &{\hat {\mib Q}}^2\ket{k-l+m,l-k+m,2s}=(m+1)((m+1)-1)\ket{k-l+m,l-k+m,2s}
\ , \nonumber\\
& &{\hat Q}_0\ket{k-l+m,l-k+m,2s}=(m+1)\ket{k-l+m,l-k+m,2s} \ . 
\end{eqnarray}
Here, the state $\ket{k-l+m,l-k+m,2s}$ is given in the relation 
(\ref{4-5}). The relation (\ref{5-18}) tells us that ${\hat Q}_+$ plays a role 
of the raising operator on the state $\ket{k-l+m,l-k+m,2s}$. 
Therefore, the eigenvalue problem for ${\hat {\mib Q}}^2$ and ${\hat Q}_0$ 
is easily solved. The case of ${\hat N}_0$ is also easily treated and the 
case of ${\hat {\mib S}}^2$, ${\hat S}_0$ and ${\hat M}_0$ is already 
investigated. Thus, we can show that the state (\ref{4-18}) is also the 
eigenstate for $({\hat {\mib S}}^2, {\hat S}_0 , {\hat {\mib Q}}^2, 
{\hat Q}_0 , {\hat M}_0 , {\hat N}_0)$ and the eigenvalues of 
${\hat {\mib Q}}^2$, ${\hat Q}_0$ and ${\hat N}_0$ are shown as follows : 
\begin{subequations}\label{5-19}
\begin{eqnarray}
& &q(q-1) \ , \ \ \quad (q=m+1=s_{12}+1) 
\qquad \hbox{\rm for} \qquad {\hat {\mib Q}}^2 
\label{5-19a}\\
& &q_0 \ , \quad (q_0=k+l+1=s_{1}+s_2+1) 
\qquad \hbox{\rm for} \qquad {\hat Q}_0 
\label{5-19b}\\
& &n_0 \ , \qquad\quad\ \quad (n_0=l-k+s=s_{3}) 
\qquad\ \hbox{\rm for} \qquad {\hat N}_0 
\label{5-19c}
\end{eqnarray}
\end{subequations}
Here, $q$, $q_0$ and $n_0$ take the values 
\begin{equation}\label{5-20}
q=1, 3/2, 2, \cdots , \qquad q_0=q, q+1, q+2, \cdots , \qquad
n_0=0, 1/2, 1, \cdots . 
\end{equation}
The relation (\ref{5-19}) gives us 
\begin{equation}\label{5-21}
m=q-1\ , \qquad k=(1/2)\cdot(q_0-n_0+s-1) \ , \qquad 
l=(1/2)\cdot (q_0+n_0-s-1) \ .
\end{equation}
Then, the state (\ref{4-18}) can be rewritten in the form 
\begin{eqnarray}
& &\ket{q, q_0,m_0,n_0;s,s_0}
=\left(\sqrt{\Gamma}\right)^{-1}({\hat S}_+)^{s+s_0}
({\hat M}_+)^{q+m_0-1}({\hat Q}_+)^{q_0-q} \nonumber\\
& &\qquad\qquad\qquad\qquad\qquad
\times ({\hat R}_+(2))^{q-n_0+s-1}({\hat T}_+(2))^{q+n_0-s-1}
({\hat b}_3^*)^{2s} \ket{0} \ . 
\label{5-22}\\
& &\Gamma=
[(2s)!(s+s_0)!/(s-s_0)!][(2q-2)!(q+m_0-1)!/(q-m_0-1)!]
\nonumber\\
& &\qquad
\times [(2s)!(q_0-q)!(n_0+q-s-1)!/(n_0-q+s+1)!]^2 
\nonumber\\
& &\qquad
\times \sum_{i=0}^{q_0-q}\sum_{j=0}^{n_0-s+q-1}
[(q_0-n_0+s-1-i+j)!(n_0+q-s-1+i-j)!(n_0+s-j)!] \nonumber\\
& &\qquad\qquad\qquad\qquad\qquad \times
[i!j!(q_0-q-i)!((n_0+q-s-1-j)!)^2]^{-1} \ .
\label{5-23}
\end{eqnarray}
The above is the fourth aspect of the orthogonal set given in \S 3.

\section{Concluding remarks}

Finally, as the concluding remarks, we contact with the relation of 
the present formalism with the $su(2,1)$-boson model in three kinds of 
boson operators, which was proposed by the present authors.\cite{6} 
In the original notations, this $su(2,1)$-model can be expressed in terms 
of the bosons $({\hat a}, {\hat a}^*)$, $({\hat c}, {\hat c}^*)$ and 
$({\hat d},{\hat d}^*)$. We make these bosons correspond to 
$({\hat a}_1, {\hat a}_1^*)$, $({\hat b}_2, {\hat b}_2^*)$ and 
$({\hat b}_3, {\hat b}_3^*)$, respectively, in the present notations. 
Further, this model consists of nine (essentially eight) 
$su(2,1)$-generators, which are denoted by ${\hat S}_{\pm,0}$, 
${\hat T}_{\pm,0}$ and ${\hat R}_{\pm,0}$. In order to avoid the confusion 
of the notations, in this paper, the above three sets are expressed as 
${\hat \sigma}_{\pm,0}(1)$, ${\hat \tau}_{\pm,0}(1)$ and 
${\hat \rho}_{\pm,0}(1)$, respectively. Then, these generators can 
be expressed as 
\begin{subequations}\label{6-1}
\begin{eqnarray}
& &{\hat \sigma}_+(1)
={\hat b}_2^*{\hat b}_3 \ , \qquad 
{\hat \sigma}_-(1)={\hat b}_3^*{\hat b}_2 \ , \qquad
{\hat \sigma}_0(1)
=(1/2)\cdot({\hat b}_2^*{\hat b}_2-{\hat b}_3^*{\hat b}_3) \ , 
\label{6-1a}\\
& &{\hat \tau}_+(1)
={\hat a}_1^*{\hat b}_2^* \ , \quad 
{\hat \tau}_-(1)={\hat b}_2{\hat a}_1 \ , \quad
{\hat \tau}_0(1)
=(1/2)\cdot({\hat a}_1^*{\hat a}_1+{\hat b}_2^*{\hat b}_2)+1/2 \ , \quad
\label{6-1b}\\
& &{\hat \rho}_+(1)
={\hat a}_1^*{\hat b}_3^* \ , \quad 
{\hat \rho}_-(1)={\hat b}_3{\hat a}_1 \ , \quad
{\hat \rho}_0(1)
=(1/2)\cdot({\hat a}_1^*{\hat a}_1+{\hat b}_3^*{\hat b}_3)+1/2 \ . \quad
\label{6-1c}
\end{eqnarray}
\end{subequations}
Here, the relation ${\hat \sigma}_0(1)={\hat \tau}_0(1)-{\hat \rho}_0(1)$ 
should be noted. Next, we replace $({\hat a}_1 , {\hat a}_1^*)$, 
$({\hat b}_2 , {\hat b}_2^*)$ and $({\hat b}_3, {\hat b}_3^*)$ in the 
forms (\ref{6-1}) with $({\hat b}_1 , {\hat b}_1^*)$, 
$(-{\hat a}_2 , -{\hat a}_2^*)$ and $(-{\hat a}_3, -{\hat a}_3^*)$, 
respectively, and another set of the $su(2,1)$-generators denoted by 
${\hat \sigma}_{\pm,0}(2)$, ${\hat \tau}_{\pm,0}(2)$ and 
${\hat \rho}_{\pm,0}(2)$ is obtained : 
\begin{subequations}\label{6-2}
\begin{eqnarray}
& &{\hat \sigma}_+(2)
={\hat a}_2^*{\hat a}_3 \ , \qquad 
{\hat \sigma}_-(2)={\hat a}_3^*{\hat a}_2 \ , \qquad
{\hat \sigma}_0(2)
=(1/2)\!\cdot\!({\hat a}_2^*{\hat a}_2-{\hat a}_3^*{\hat a}_3) \ , 
\label{6-2a}\\
& &{\hat \tau}_+(2)
=-{\hat b}_1^*{\hat a}_2^* \ , \quad 
{\hat \tau}_-(2)=-{\hat a}_2{\hat b}_1 \ , \quad
{\hat \tau}_0(2)
=(1/2)\!\cdot\!({\hat b}_1^*{\hat b}_1+{\hat a}_2^*{\hat a}_2)+1/2 \ , \qquad
\label{6-2b}\\
& &{\hat \rho}_+(2)
=-{\hat b}_1^*{\hat a}_3^* \ , \quad 
{\hat \rho}_-(2)=-{\hat a}_3{\hat b}_1 \ , \quad
{\hat \rho}_0(2)
=(1/2)\!\cdot\!({\hat b}_1^*{\hat b}_1+{\hat a}_3^*{\hat a}_3)+1/2 \ . 
\quad\qquad
\label{6-2c}
\end{eqnarray}
\end{subequations}
If the above set is added to that shown in the relation (\ref{6-1}), 
we obtain new set expressed as follows : 
\begin{subequations}\label{6-3}
\begin{eqnarray}
& &{\hat \sigma}_\pm={\hat \sigma}_\pm(1)+{\hat \sigma}_\pm(2)
={\hat R}_{\pm}(2) \ , \qquad
{\hat \sigma}_0={\hat \sigma}_0(1)+{\hat \sigma}_0(2)
={\hat R}_2^2 \ , 
\label{6-3a}\\
& &{\hat \tau}_\pm={\hat \tau}_\pm(1)+{\hat \tau}_\pm(2)
={\hat Q}_{\pm} \ , \qquad
{\hat \tau}_0={\hat \tau}_0(1)+{\hat \tau}_0(2)
={\hat Q}_0 \ , 
\label{6-3b}\\
& &{\hat \rho}_\pm={\hat \rho}_\pm(1)+{\hat \rho}_\pm(2)
={\hat T}_{\pm}(1) \ , \qquad
{\hat \rho}_0={\hat \rho}_0(1)+{\hat \rho}_0(2)
={\hat T}_1^1 \ . 
\label{6-3c}\\
\end{eqnarray}
\end{subequations}
Translation of an orthogonal set given in Ref. \citen{6} to the present 
system leads us to the form 
\begin{subequations}\label{6-4}
\begin{eqnarray}
\ket{\kappa,t,t_0}&=&
({\hat \tau}_+)^{t_0-t}({\hat \sigma}_+)^{2t-1}({\hat b}_3^*)^{2\kappa-2}
\ket{0} \nonumber\\
&=&({\hat Q}_+)^{t_0-t}({\hat R}_+(2))^{2t-1}({\hat b}_3^*)^{2\kappa-2}
\ket{0} \ , 
\label{6-4x}\\
\kappa&=&1, 1/2, 2, \cdots \ , \quad 
t=1/2, 1, 3/2, \cdots \ , \quad
t_0=t, t+1, t+2, \cdots \ . \qquad
\label{6-4a}
\end{eqnarray}
\end{subequations}
Comparison of the state (\ref{6-4}) with the form (\ref{4-18}) gives us 
\begin{eqnarray}
& &s_0=-s \ , \qquad m_0=-m \ , \qquad m=k-l \ , 
\label{6-5}\\
& &\kappa=s+1\ , \qquad t=k-l+1/2 \ . \qquad t_0=k+l+1/2 \ . 
\label{6-6}
\end{eqnarray}
The above is the relation of the formalism developed in this paper to the 
$su(2,1)$-boson model presented by the present authors. 
Of course, it should be noted that the number of the degrees of freedom 
increases twofold.

As was mentioned in \S 1, the aim of the present paper is to give a 
preparation for formulating the deformed boson scheme in six kinds of 
boson operators. For this aim, the starting point may be seen in the 
following coherent state : 
\begin{eqnarray}\label{6-7}
\ket{c^0}&=&\left(\sqrt{\Gamma_c}\right)^{-1}
\exp \left(\gamma_S{\hat S}_+\right) \cdot \exp\left(\gamma_M{\hat M}_+
\right) \nonumber\\
& &\times \exp\left(\gamma_Q{\hat Q}_+\right)\cdot
\exp\left(\gamma_T{\hat T}_+(2)\right)\cdot
\exp\left(\gamma_R{\hat R}_+(2)\right)\cdot
\exp\left(\delta{\hat b}_3^*\right)\ket{0} \ . 
\end{eqnarray}
The symbols $\gamma_S$, $\gamma_M$, $\gamma_Q$, $\gamma_T$, $\gamma_R$ and 
$\delta$ denote complex parameters. In Ref. \citen{12}, 
a special case $\gamma_S=\gamma_M=\gamma_T=0$ was treated. 
It may be interesting 
to investigate the deformation of the coherent state $\ket{c^0}$ based 
on the deformed boson scheme which has been developed by the present 
authors. In subsequent paper, we will discuss this problem.

%\section*{Acknowledgements}

\end{document}